\documentclass[12pt]{iopart}
\input{epsf.sty}
\usepackage{multicol}
\usepackage{epsfig,bbm,bm}
\usepackage{amssymb}

\def\be{\begin{equation}}
\def\ee{\end{equation}}    
\def\baray{\begin{eqnarray}}
\def\earay{\end{eqnarray}}
\def\dper{d_\perp}
\def\ba{\begin{eqnarray}}
\def\ea{\end{eqnarray}}

\def\rpl{r_\parallel}
\def\Vp{V_\perp}
\def\Vpl{V_\parallel}

\def\lpl{\ell_\parallel}

\def\2pi{\left(2\pi\right)}

\def\hf{\frac{1}{2}}
\def\DD{D \bar D}
\def\Vol{{\mathcal V}}
\def\hf{\frac{1}{2}}

\begin{document}

\title[Inter-brane Interactions]{Inter-brane Interactions in Compact Spaces and Brane Inflation}

\author{Sarah Shandera\dag\ Benjamin Shlaer\dag\ Horace Stoica\ddag\ and S.-H. Henry Tye\dag}

\address{\dag\ Laboratory for Elementary Particle Physics\\ 
Cornell University\\ 
  Ithaca, NY 14853}
  \ead{seb56@cornell.edu}
  \ead{shlaer@lepp.cornell.edu}
  \ead{tye@lepp.cornell.edu}

\address{\ddag\ Ernest Rutherford Physics Building\\
McGill University\\
3600 rue University\\
Montr\'eal, QC\\
Canada H3A 2T8}

\ead{stoica@physics.mcgill.ca}

\begin{abstract}
 It was pointed out that brane-anti-brane inflation without 
 warped geometry is not viable due to compactification effects 
 (in the simplified scenario where the inflaton is decoupled from the 
 compactification moduli).  
 We show that the inflationary scenario with branes at a small angle 
 in this simplified scenario remains viable. 
 We also point out that brane-anti-brane inflation
 may still be viable under some special conditions.
 We also discuss a way to treat potentials in compact spaces that 
 should be useful in the analysis of more realistic brane inflationary 
 scenarios.
\end{abstract}

%\pacs{Brane world, Inflation, Jellium term, Compact Space}
\submitto{JCAP}
\maketitle
\section{Introduction}

Brane interactions in the brane world have been proposed as 
the origin of inflation \cite{Guth:1980zm,Linde:1981mu,Albrecht:1982wi}, 
an epoch in the early universe that 
initiated the radiation-dominated big bang. 
In brane inflation \cite{Dvali:1998pa} 
the inflaton is an open string mode, while the brane interaction 
comes from 
the exchange of closed string modes. A particularly simple 
scenario is the $Dp$-brane-anti-$Dp$-brane system. Although 
the potential between them seems too steep, it was proposed 
\cite{Burgess:2001fx} that the compactification 
effect in the special case of a hyper-cubic torus will flatten
the inflaton potential for enough inflation. Let us call this 
the $\DD$ scenario.

Recently, it was pointed out \cite{Maldacena,Kachru:2003sx} that 
the Poisson equation for the inflaton potential $\Phi$ in a 
compactified manifold should include a background term
(the so-called ``jellium'' term in solid state physics).
As a result, the slow-roll parameter $\eta$ in the $\DD$ 
inflationary scenario (in the simplified version where 
the stabilization of compactification moduli is independent 
of the inflaton) becomes
\be
\label{mald}
\eta \simeq -2/\dper
\ee
where $\dper$ is the number of dimensions perpendicular to the 
branes, $\dper = 9-p \le 6$. 
Since we need at least $N_e=50$ 
e-folds of inflation, and $|\eta| < 1/N_e$, the $\DD$ scenario
is not viable as an inflationary model.
The impact of the jellium term is clearly important to the 
analysis of the inflationary properties in some of these 
scenarios. This was first studied in Ref.\cite{Rabadan:2002wy}.

Here we would like to point out that the brane inflationary 
scenario where branes are at a small angle 
\cite{Garcia-Bellido:2001ky,Jones:2002cv} 
remains robust because the jellium term
is much smaller in this scenario. In the $\left(n, 1\right)$, 
$\left(n, -1\right)$ wrapping scenario (which reduces to $2n$ 
parallel D4-branes after inflation), 
\be
\eta \simeq -\frac{\theta^2}{4n}
\ee
where $\theta$ is the angle between the two branes 
($\theta \simeq 1/12$ and $n=8$ are reasonable values).
When the jellium contribution to $|\eta|$ is much less than $1/50$, the 
slow-roll behavior of the inflaton is dictated by the other 
terms in the potential, in particular the quartic harmonic 
term as measured around the antipodal point, as studied in 
Ref.\cite{Jones:2002cv}.

We would also like to point out that the $\DD$ scenario may 
still be possible under some special conditions. Whether that 
special condition on the background charge distribution is
realized or not depends on the dynamics of moduli stabilization 
(the dilaton, the complex and K\"{a}hler structures 
of the compactified manifold etc.), an issue that needs 
better understanding \cite{Giddings:2001yu,Kachru:2003aw,Kachru:2003sx}.

In section 2, we give a review of the Poisson equation in compact 
spaces. A simple ansatz of calculating the potential energy between 
two charges is $q_1 \Phi_2$ where $\Phi_2$ is the potential 
due to the the second charge $q_2$, irrespective of whether the source 
charges in compact space add to zero or not. We shall justify this 
ansatz by showing that it is equivalent (up to a constant) to
the potential energy between two charges (NS-NS or RR) by 
directly integrating the potential energy density over the compact 
space. This equivalence is simply illustrated in the 1-dimensional
case \cite{Rabadan:2002wy}. In higher dimensions, the Green's function 
in compact space may be represented in terms of Jacobi theta functions. 
In section 4, we generalize the method 
used in solid state physics to show how to obtain the Green's function
numerically. This method should be useful in more realistic brane world
models. For flat compact spaces, the Green's functions obtained 
by these two methods agree. 
In section 5, we apply the result to brane inflation in the
cosmological context. As pointed out in Ref.\cite{Maldacena,Kachru:2003sx},
the slow-roll parameter $\eta$ is too big for the brane-anti-brane 
scenario. On the other hand, the impact of the jellium term in the 
branes-at-small-angle scenario can be made negligibly small. In this 
sense, the branes-at-small-angle inflationary scenario remains robust.
Section 6 contains discussion.

\section{Poisson Equation and Potential Energy in Compact Spaces}

In non-compact space, the determination of the 
potential energy (Coulombic or gravitational) between two charges 
is well known: we treat
one of the charge $q_1$ to be a probe charge, and the potential 
energy is given by $q_1\Phi_2$, where $\Phi_2$ is the potential 
due to charge $q_2$ (even if $|q_1|>|q_2|$). Here, we argue that
this simple ansatz is equally applicable in compact spaces, 
where $\Phi_2$ includes the 
contribution of the jellium term. That is, the potential energy
between two charged (Coulombic, Ramond-Ramond, gravitational or 
NS-NS) objects in a compact space will include the same
quadratic component due to the jellium term, irrespective of whether 
the sum of the source charges in the compact space vanishes or not. 

Consider a compact manifold $M$ ($\partial M = \emptyset$).
The Green's function is given by the Poisson equation,
\ba
\label{jeli}
\nabla^2 G({\bf r}, {\bf r^{\prime}})=
\delta\left({\bf r}-{\bf r^{\prime}} \right)- 1/\Vol
\ea
where $\delta\left({\bf r}-{\bf r^{\prime}}\right)$ is the
$\dper$-dimensional $\delta$ function and $\Vol$ is the volume of 
the compactified manifold. Integrating both sides over the 
$\dper$-dimensional compact space 
and using Stoke's theorem, we obtain Gauss's law.
Since a compact space has no boundary, this volume
integral over $\nabla^2 G$ vanishes, as does the integral over
the RHS of Eq.(\ref{jeli}).

Consider the eigenfunctions $u_{\lambda}\left({\bf r}\right)$
of the Laplacian operator on the compact manifold with 
eigenvalues $\lambda \le 0$ 

\be
\label{eigene}
\nabla^2 u_{\lambda}\left({\bf r}\right) = \lambda u_{\lambda}\left({\bf r}\right).
\ee
The eigenfunctions satisfy the completeness relation:
\be
\sum_{\lambda}u_{\lambda}\left({\bf r}\right)u_{\lambda}\left({\bf r}^{\prime}\right)
=\delta\left({\bf r}-{\bf r}^{\prime}\right)
\ee
Integrating both sides of Eq.(\ref{eigene}) over $M$, we obtain 
($dv=d^{\dper}r \sqrt{-g}$),
\be
\label{intuz}
 \int_M dv ~ u_{\lambda} ({\bf r}) = 0  \quad \quad \lambda \ne 0
\ee
In a compact space there is a normalizable zero mode of the 
Laplacian, $\nabla^2 u_{0}\left({\bf r}\right)=0$, where
$u_{0}\left({\bf r}\right)=1/\sqrt{\Vol}$. 
The Green's function can be written as:
\be
\label{Blue_Function}
G\left({\bf r},{\bf r}^{\prime}\right)=\sum_{\lambda\neq 0}\frac{u_{\lambda}\left({\bf r}\right)
u_{\lambda}\left({\bf r}^{\prime}\right)}{\lambda}
\ee
Note that the zero mode is absent in $G({\bf r},{\bf r}^{\prime})$ in 
Eq.(\ref{Blue_Function}), thus avoiding an obvious divergence. 
Using Eq.(\ref{intuz}) and (\ref{Blue_Function}), we have
\be
\label{intGz}
\int_M dv ~ G({\bf r}, {\bf r^{\prime}})=0
\ee

Let $\Phi$ be the static potential due to a given set 
of charges $q_i$ (at ${\bf r}_i$), $i=1,2,...$,
\be
\label{PhiG}
\Phi ({\bf r})= \sum_i q_iG({\bf r}, {\bf r}_i)
\ee
Using Eq.(\ref{PhiG}) and (\ref{jeli}), the total static energy due to 
this set of charges $q_k$ is
\ba
\label{tenergy}
V({\bf r}_k) = \hf \int_M dv (\nabla \Phi)^2
=-\int_M dv~ \hf \Phi\nabla^2 \Phi \\\nonumber
= - \hf \sum_i q_i \Phi({\bf r}_i) + 
\frac{\sum q_i}{2 \Vol} \int_M dv \Phi({\bf r})
\ea
Using Eq.(\ref{intGz}) and Eq.(\ref{PhiG}), we see that
the last term in Eq.(\ref{tenergy}) vanishes. 
So, irrespective of whether or not the net charge vanishes,
\be
\label{UE}
V({\bf r}_k)= - \sum_{i>j} q_iG({\bf r}_i,{\bf r}_j)q_j + constant
\ee
where the constant comes from the ${\bf r}_i$-independent 
self-interaction terms. 
When applied to the RR interaction, where $q_i=\mu_i$, we must
have $\sum_i \mu_i=0$ in our compact space. 
%The presence of the jellium term is encoded in $G({\bf r}_i,{\bf r}_j)$. 
For the NS-NS interaction, $q_i$ are the brane tensions $T_i$, $\sum_i T_i \ne 0$
and $V_{NS-NS}({\bf r}_i)$ has the opposite sign since same sign 
charges attract. Putting them together we obtain, up to a constant
\be
V({\bf r}_k)= + \sum_{i>j} (T_iT_j -\mu_i\mu_j)G({\bf r}_i,{\bf r}_j)
\ee
where the presence of the jellium term is encoded in
$G({\bf r}_i,{\bf r}_j)$. In the two brane potential, the ``Coulomb'' 
($1/r^{d-2}$)
term and the jellium ($r^2$) term have the same effective strength, namely $T_1T_2 -\mu_1\mu_2$. This effective strength is very small 
for branes at a small angle.  Since the constant term in the potential is 
linear in the brane tension $T$, $V''/V$ can be made arbitrarily small.

\subsection{One-dimensional Example}

This well-known example (given in Ref.\cite{Rabadan:2002wy}) illustrates 
the above result.
Consider a circle of circumference $L$. For $ 0 \le y \le L$, 
\be
G(y)= - \frac{y^2}{2L} + \frac{|y|}{2} -\frac{L}{12}
\ee
where the constant term may be determined by the $\zeta$-function 
regularization of the divergent Green's function $\hat G$ of 
the 1-dimensional lattice, 
\be
2 \hat G(y) = |y| + \sum_{n=1}^{\infty} |y+nL| +|y-nL|
\ee
Notice that $G(y)$ satisfies Eq.(\ref{intGz}).

Let us put a charge $q_1=q$ at $y_1=0$ and a second charge $q_2=-q$ 
at $y_2=y$,
where $0 \le y < L$ so that the total charge in the circle is zero. 
Using Eq.(\ref{UE}), one finds 
\be
\label{UE1}
V(y)= q^2G(y) + \frac{q^2L}{12} = 
q^2 \left( \frac{|y|}{2} -\frac{y^2}{2L} -\frac{L}{12} \right) 
+\frac{q^2L}{12}
\ee
where the first term is the linear potential and the quadratic term is 
due to the jellium effect. The last term is the self-energy term
and is independent of $y$. As we put the charges on top of each other,
$V(0)=0$, as expected. 

We may also calculate the potential at $x$ due to these two charges
($+q$ and $-q$) :
\ba
\phi_1 (x) &=& q \frac{x}{L}(L-y) -qy +\phi_0  \quad  \quad 
0 \le x \le y \\\nonumber
\phi_2(x) &=& -\frac{qxy}{L} +\phi_0 \quad  \quad  y \le x \le L
\ea
Note that $\phi(x)$ is continuous and piecewise linear in $x$, 
i.e., $\phi_1(0)=\phi_2(L)$ and $\phi_1(y)=\phi_2(y)$.
Now we can integrate over the circle the energy density 
($(\nabla \phi)^2/2$) to obtain the energy as a function of 
the relative position $y$ of the charges.
This exactly reproduces $V(y)$ in Eq.(\ref{UE1}). 
In short, $\phi(x,y)$ is the potential for a probe charge at $x$, 
while $V(y)$ is what we are interested in.

For the case of two masses on a circle: $m_1$ at $y_1=0$ and 
$m_2$ at $y_2=y$, the potential at $x$ ($0 \le x < L$) due 
to these 2 masses is
$\phi(x) = m_1 G(x) + m_2 G(x-y)$, where there is a jellium 
contribution in each Green's function. From this, 

\ba
\nabla\phi_1 (x) = -\frac{m_1 + m_2}{L}x + \frac{m_2y}{L} + \frac{m_1 - m_2}{2}  \quad  \quad 0 \le x \le y \\\nonumber
\nabla\phi_2(x) =  -\frac{m_1 + m_2}{L}x + \frac{m_2y}{L} + \frac{m_1 + m_2}{2}  \quad  \quad  y \le x \le L
\ea

and, as expected, we have
\be
V(y) =  - \hf \int dx (\nabla \phi)^2
=m_1 m_2 G(y) - \frac{(m_1^2+m_2^2)L}{24}    \nonumber
\ee
where the last constant term is the self-energy contribution.

\section{Green's Function in Arbitrary Dimension}

When two branes are moving in a compactified space, we need to 
calculate the inter-brane potential. 
This requires finding $\Phi({\bf r})=-G({\bf r})$
that satisfies Eq.(\ref{jeli}). 
Our main goal is to determine the Green's 
function around the antipodal point (the point farthest from the source). 
To simplify the problem, consider a square, flat $d$-dimensional torus of volume 
$L^d$. Using the identity
\be
\frac{1}{L^2\lambda} = -\int_0^\infty e^{L^2\lambda s}ds \quad \quad (\lambda < 0)
\ee
and the eigensolutions
\baray
u_{\bf n}({\bf x}) &=& \sqrt{L^{-d}}e^{2\pi i {\bf n}\cdot {\bf x}/L}, \quad \quad \lambda_{\bf n} = -\frac{4\pi^2n^2}{L^2}
\earay
we can regularize
\baray
\label{theta}
G({\bf x}) &=& -\sum_{{\bf n}\in {\mathbbm Z}^d}(1 - \delta_{{\bf n},{\bf 0}})\frac{e^{2\pi i {\bf n}\cdot {\bf x}/L}}{4\pi^2n^2L^{d-2}}\\\nonumber
&=& L^{2-d}\int_0^\infty ( 1-\prod_{j=1}^d\sum_{n_j\in{\mathbbm Z}}e^{2\pi in_jx_j/L-4\pi^2 n_j^2s})ds\\\nonumber
&=& L^{2-d}\int_0^\infty \left[1 - \prod_{j = 1}^d\theta_3(\frac{\pi x_j}{L}, e^{-4\pi^2s})\right]ds
\earay
Where $\theta_3$ is a Jacobi theta function \cite{Whittaker}
\be
\theta_3\left(\frac{b}{2},\e^{-a}\right)=\sum_n\e^{-an^2+ibn}
\ee
The rapid convergence of the above expression makes the flat $d$-torus especially simple to numerically simulate.

When one considers compactification on a manifold ($T^2$, $K3$, $CY_3$) with
a non-trivial metric, another numerical method suggests itself.  Developed
by Ewald and others \cite{ewald}, this method is suited to spaces that
may be represented as a periodic lattice (perhaps modulo some additional 
discrete symmetries). To illustrate the method, we proceed
with a simple example, a square, flat torus. Let us again write 
$\Phi$ as in Eq.(\ref{Blue_Function}): 
\be
\label{Ewald}
\Phi({\bf r}) = \frac{1}{\Vol}\sum_{{\bf k}_j\neq {\bf 0}}
\frac{\exp(i{\bf k}_{j}\cdot{\bf r})}{k_{j}^2} \quad
\ee
where ${{\bf k}_j}\in \frac{2\pi}{L}{\mathbbm Z}^d$ is a reciprocal lattice vector and $k_j$ 
its magnitude. 
This sum diverges for $d \ge 2$, since the number of terms grows as $k_j^{d-1}$ (i.e., with the 
surface area of a $d$-dimensional ball).  Alternatively, one may treat the torus as an infinite lattice and sum over the source and its images at the lattice points:
\be
\Phi({\bf r}) = \quad \sum_{{\bf r}_j}
\frac{\alpha}{({\bf r}-{\bf r_j})^{d-2}}
\ee
where
\be
\label{alph}
\alpha=\frac{\Gamma(\frac{d-2}{2})}{4\pi^{\frac{d}{2}}}, \quad d\neq2
\ee
 As we saw in the one-dimensional case, this will lead to a divergence and the need to regularize it. The answer depends on how the lattice points are summed and how the regularization is carried out. 

Ewald's method is to add and 
subtract a diffuse charge distribution around each lattice point. 
The added 
charge makes the monopole moment of each cell zero, 
aiding the convergence of
the real lattice sum. 
This additional charge (minus a jellium term) is convergently subtracted away 
in reciprocal space, since a reciprocal space sum converges for non-singular charge distribution. 
The final result is independent of the diffuse charge
distribution used to regulate the sums. Details on how to implement this method in two, three and four dimensions are included in the appendix. The numerical data generated 
by the Ewald method may be fit to an expression for the potential near the source 
of the form:
\ba
\label{genform}
\Phi^{(d)}({\bf r})&=& \frac{\alpha}{r^{d-2}} + \frac{r^2}{2d\Vol} + \Phi^{(d)}_{harmonic}({\bf r}) + constant, \quad d>2\\\nonumber
&=&-\frac{1}{2\pi}\ln\left(\frac{r}{L}\right) + \frac{r^2}{4\Vol} + 
\Phi^{(2)}_{harmonic}({\bf r}) + constant , \quad d=2
\ea
Notice that although individual terms in 
$\Phi^{(d)}({\bf r})$ are not periodic, their sum is. 
The constant is fixed by requiring $\Phi^{(d)}({\bf r})$ to be 
independent of the added and subtracted charge.  
The harmonic piece satisfies the Laplace equation 
(when restricted to the cell containing the origin):
\be
\label{Laplace}
\nabla^2\Phi^{(d)}_{harm}=0
\ee
For a hypercubic torus (with $\Vol=L^d$) it has the general form
\be
\label{genform2}
\Phi^{(d)}_{harm}({\bf r)}= \frac{C^{(d)}_{s}}{\Vol^{\frac{d-2}{d}}} 
+ A^{(d)}_4h^{(d)}_4({\bf r})+ A^{(d)}_6h^{(d)}_6({\bf r})+\dots
\ee
where the $h^{(d)}_{n}$ are polynomials of order $n$ with coefficients 
determined by Eq(\ref{Laplace}). For example, for a hyper-cubic lattice 
with coordinates $x_i$ measured from the source (podal point) and ${\bf r}= (x_1,x_2,...,x_d)$
\ba
h^{(d)}_4({\bf r}) &=& \frac{1}{\Vol^{\frac{d+2}{d}}}\left[\sum_{i=1}^{d}x_i^4-\frac{6}{d-1}\sum_{i\neq j}x_i^2x_j^2\right]\\\nonumber
h^{(d)}_6({\bf r}) &=& \frac{1}{\Vol^{\frac{d+4}{d}}}\left[\sum_{i\neq j\neq k}x_i^2x_j^2x_k^2+\frac{(d-2)(d-1)}{180}\sum_{i}x_i^6
-\frac{(d-2)}{12}\sum_{i\neq j}x_i^4x_j^2\right]
\ea
In two dimensions, terms of order $4m + 2$ are not 
present in the harmonic piece. Beyond the sixth order terms in the 
hypercubic case, and for $d>2$ for a rectangular lattice, there is more 
than one undetermined coefficient $A^{(d)}_{n,i}$ at each order. 
A two-dimensional rectangular lattice has only one parameter at any 
even order. At a given order $n$, the number of parameters 
for a hypercubic lattice reaches a maximum and becomes independent of dimension for $d \geq \frac{n}{2}$.

There is an expression similar to Eq.(\ref{genform}) for the potential 
near the antipodal point. This is the expression that is suitable 
for applications in brane inflation. 
With coordinates $z_i$ now measured from the 
antipodal point (${\bf z}= {\bf r} - (L/2,L/2,...)$), 
$\Phi$ has the form
\be
\label{antipodal}
\Phi^{(d)}_{antipodal}({\bf z})= \frac{C^{(d)}_{a}}{\Vol^{\frac{d-2}{d}}} + \frac{1}{2d\Vol}\sum_{i=1}^dz_i^2 + B^{(d)}_4h^{(d)}_4({\bf z})+ B^{(d)}_6h^{(d)}_6({\bf z})+\dots
\ee
The results for the coefficients in Eq.(\ref{genform}) and 
Eq.(\ref{antipodal}) were obtained by the method described below, 
and are summarized in Tables \ref{sourcecoef} and 
\ref{antipodalcoef}. The lattice spacing has been set to one. 
At least four terms in $\Phi_{harm}$ were kept in each dimension, but 
the accuracy of the numerical values could be improved by keeping more. 
In general, the convergence of Eq.(\ref{genform}) is somewhat better 
than that of Eq.(\ref{antipodal}). 

It is easy to check that the integral of $\Phi$ over a unit cell is zero,
that is, $\Phi$ satisfies Eq.(\ref{intGz}). 
In solid state physics, the Madelung constant is found by considering the potential due to both the positive and negative ions. Since the negative ions are found at the antipodal point in a simple cubic lattice, the Madelung constant is given by $\alpha_m = C_s-C_a$. In three dimensions, our value for the Madelung constant of a simple cubic lattice agrees with Ref.\cite{kittel}. The constants $C_a$ in Table \ref{antipodalcoef} agree with
Eq.(\ref{theta}) evaluated at the antipodal point,
and are four times those listed in Ref.\cite{Burgess:2002}.

For rectangular torus, there will be quadratic harmonic terms of the form
$z_i^2-z_j^2$. Their impact on inflation is discussed in 
Ref.\cite{Jones:2002cv}.
The hypercubic way to sum the lattice generates only the harmonic 
terms. The numerical values of $B_4$ is at least a factor of 3 
smaller than that given in Ref.\cite{Jones:2002cv}. 
This weakens the potential and improves the inflationary
scenario. 

\begin{table}
\caption{\label{sourcecoef}Constant, fourth order and sixth order 
coefficients in potential near source.}
\begin{center}
\begin{tabular}{@{}cccc}
\br
 $\dper$& 2& 3& 4\\
\hline
$C_{s}$ & -0.21& -0.21& -0.17\\
\hline
$A_4$ & 0.12& 0.44& 0.34\\
\hline
$A_6$ & 0.00& 0.0072& 3.05\\
\hline
\end{tabular}
\end{center}
\end{table}

\begin{table}
\caption{Constant and coefficients of the fourth order and sixth order 
terms in the potential near the antipodal point \label{antipodalcoef}.}
\begin{center}
\begin{tabular}{@{}cccc}
\br
 $\dper$& 2& 3& 4\\
\hline
$C_{a}$ & -0.055& -0.064& -0.070\\
\hline
$B_4$ & -0.62& -0.53& -2.20\\
\hline
$B_6$ & 0.00& 0.0024& -101.5\\
\hline
\end{tabular}
\end{center}
\end{table}

\section{Application to Brane Inflationary Scenarios}

Let us consider a few brane inflationary scenarios, where the moduli 
stabilization effects are ignored.
In order to find the potential
of the inflaton as seen by a 4D observer we need to calculate the 
low-energy effective action for the brane system. The interaction between 
the branes due to the exchange of closed strings 
depends on their separation, so we will decompose the coordinates of 
the two (stacks of) branes into the center-of-mass and the relative 
separation.  

Assuming that the branes wrap $n$ and $m$ times the volume $\Vpl$, the low-energy
effective action is obtained by expanding the DBI action:
\baray
&& S_{eff} = n\tau_p\int d^{p+1}\xi_1 \sqrt{1+\partial_{\mu}y_1\partial^{\mu}y_1} + 
m\tau_p\int d^{p+1}\xi_2 \sqrt{1+\partial_{\mu}y_2\partial^{\mu}y_2} \nonumber \\
&& \simeq \left(m+n\right)\lpl\tau_p  + 
n\tau_p\int d^{p+1}\xi_1 \hf\partial_{\mu}y_1\partial^{\mu}y_1 + 
m\tau_p\int d^{p+1}\xi_2 \hf\partial_{\mu}y_2\partial^{\mu}y_2 
\earay 
where $\tau_p$ is the brane tension
\be
\tau_p= M_s^{p+1}/(2 \pi)^p g_s
\ee
where $M_s$ is the string scale and $g_s$ the string coupling.
The coordinates of the brane in the transverse directions are expressed as:
\baray
y_1 &=& y_{CM} + \frac{m}{m+n}y_r \\
y_2 &=& y_{CM} - \frac{n}{m+n}y_r 
\earay
and substituting these into the expression for $S_{eff}$ we obtain:
\be
S_{eff} = \tau_p\frac{mn}{2\left(m+n\right)}\int d^{p-3}\xi\int d^4\xi \partial_{\mu}y_r\partial^{\mu}y_r = 
\int d^4\xi \hf\partial_{\mu}\psi\partial^{\mu}\psi
\ee
The relationship between $y_r$ and $\psi$ is given by:
\be
\label{field_redef}
\psi = y_r\sqrt{\frac{mn}{\left(m+n\right)}\tau_p\Vpl}
\ee
where $\Vpl=\int d^{p-3}\xi=\lpl^{p-3}$.

\subsection{The $\DD$ scenario}

The $\DD$ potential is given in Ref.\cite{Burgess:2001fx,Jones:2002cv},
with $\dper=9-p$ :

\be
V\left(y\right)=2\tau_p\lpl^{p-3} - 
\frac{\kappa^2\beta\tau_p^2\lpl^{p-3}}{y^{\dper-2}}
\ee
where 
\be 
\kappa^2 = 8\pi G_{10} = \frac{g_s^2\left(2\pi\right)^7}{2M_s^8}
\ee
and $\beta = 2 \alpha$ given in Eq.(\ref{alph}).
Measured relative to the antipodal point, the position of the 
anti-brane is given by ${\bf z}$, which is simply ${\bf y}$ shifted 
by half the lattice size. 
The important pieces of the potential for the slow-roll condition 
on inflation 
are the constant term and the quadratic piece due to the jellium term : 
%Including the effect of the background charge, the potential is given by:
\be
\label{brane_anti_brane_quadratic}
V\left(z\right)=2\tau_p\lpl^{p-3}- 
z^2\frac{\kappa^2\tau_p^2\lpl^{p-3}}{\dper\Vp}
\ee
where $\Vol=\Vp$. As pointed out in 
Ref.\cite{Maldacena,Kachru:2003sx}, the relevant slow-roll parameter 
$\eta$ is given by:
\be
\eta=M_P^2 \frac{V^{\prime\prime}}{V} \simeq - \frac{2}{\dper}
\ee
where the derivative is taken with respect to the scalar field $\psi$ that 
appears in the low-energy effective theory, Eq.(\ref{field_redef}).
Since $\dper \le 6$, the slow-roll condition is never satisfied 
in this case, 
that is, the branes will collide far too early for any significant 
inflation to take place. 

A priori, it is still possible that the stabilization dynamics
of the extra dimensions has some unusual features that suppress 
$\eta$ and realize
the condition required for the viability of the $\DD$ inflationary
scenario. One such possibility utilizes a warped geometry to suppress 
the inter-brane attractive potential \cite{Kachru:2003sx}. 

Here, let us consider 
\ba
\nabla^2\Phi &=& \sum_i q_i\delta\left({\bf r}-{\bf r}_{i}\right) - 
F({\bf r}- (L/2, L/2,..)) \\\nonumber
F({\bf z}) &=& \frac{\sum_i q_i}{\Vol} + f({\bf z})
\ea
where $f({\bf z})$ is multi-periodic and consistency requires it to 
satisfy
\be
\int_M dv f({\bf z}) = 0
\ee
It is not hard to imagine that $f({\bf z})$ originates from the
stabilization of the moduli in the extra dimensions.
As one can easily see, $\DD$ inflationary scenario is viable if
$F({\bf z})$ vanishes at the antipodal point. 
To suppress the quadratic term in the inflaton potential, we 
need to decrease the value of $F({\bf z})$ at 
the antipodal point so that
\be
|\eta|  \simeq  |2 \Vol F({\bf 0})/\dper| < 1/N_e 
\ee 
Suppose, for a torus, measured with respect to the antipodal point, 
\be
F({\bf z}) =\frac{1-\Pi\cos(k_jz_j)}{\Vol} \simeq 
\frac{\sum (k_jz_j)^2}{\Vol}
\ee
so that $F({\bf 0})=0$. This implies that the inflaton potential 
does not have an anharmonic quadratic term around the antipodal point. 
Such a $\DD$ scenario will be able to give enough inflation to 
render the model viable. (In fact, one may choose $F({\bf z})$
to reduce the contribution of the quartic term to $\eta$ as well.)
Among other factors, the form of $F({\bf z})$ 
depends on the dynamics of moduli stabilization, an issue that is 
not fully understood\cite{Giddings:2001yu,Kachru:2003aw,Kachru:2003sx}. 
Generically, we should consider $F({\bf 0}) << 1/\Vol$ 
as a fine-tuning. Since $\eta$ is generically around 1 for the
$\DD$ system, we need a fine-tuning of 1 in 100 on 
$F({\bf z})$ to suppress $\eta$. 
In more realistic constructions 
of string models, it will be very interesting to see how such a 
condition can be satisfied.

\subsection{Branes at a Small Angle}

Let us consider the simple scenario \cite{Garcia-Bellido:2001ky}
where $p = 4$, $\dper = 4$, 
but with different wrappings of the branes. The $X^4, X^5$ torus, has 
sides $\lpl$ and $u\lpl$ and the branes wrap one-cycles on the torus, 
as shown in Figure 1. These branes are separated in the $X^6, X^7,
X^8, X^9$ directions by a distance $y$.

\begin{figure}[htb]
\begin{center}
\includegraphics[width=0.5\textwidth,angle=0]{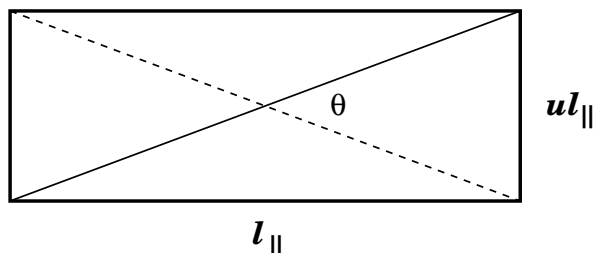} 
\caption{Two branes wrapping different cycles in a rectangular torus. The
angle between the branes is $\tan\theta\simeq\theta=2u$.
\label{torus_wrapping}}
\end{center}
\end{figure}

The Planck mass is given by 
\be
M_P^2 = \frac{M_S^8\lpl^2 u \Vp}{g_s^2\pi\left(2\pi\right)^6}
\ee
We start with the example shown in Figure \ref{torus_wrapping} where 
the wrapping numbers are $\left(1, 1\right)$ and $\left(1, -1\right)$.
We shall consider small $\theta$,  so the 
angle between the branes is $\theta\simeq\tan\theta=2u$. 
The constant piece of the potential is given by:
\be
V_0=\tau_p\lpl\left(2\sqrt{1+u^2}-2\right)\simeq\tau_p\lpl u^2
\ee
and the full potential is:
\be
\label{quadratic_potential}
 V\left(y\right) = \tau_{4}\lpl u^2 - \frac{Q y^2}{2\dper\Vp} - \frac{\beta Q}{2y^2} + harmonic 
\ee 
where $Q$ includes the contributions from both the NS-NS and RR sectors.
\be
Q = \frac{M_s^2}{2 \pi \sin \theta}
\left(1 -\frac{\sin^2 \theta}{2}- \cos \theta \right)
\simeq M_s^2\theta^3/16\pi \simeq M_s^2u^3/2\pi 
\ee 
where the $(1 - \sin^2 \theta/2)$ term comes from the NS-NS sector 
while the $-\cos \theta$ term comes from the RR sector. The $\sin \theta$
in the denominator comes from the length of the brane along the 
$\lpl$ direction. ${\bf y}$ measures the relative positions of the branes.
With the above expressions
for the potential and the Planck mass, we evaluate the potential 
$V({\bf z})$ at the antipodal point.  The relevant slow-roll parameter 
$\eta$ there becomes:
\be
\eta=M_P^2\frac{V^{\prime\prime}}{V}= 
-\frac{M_S^8\lpl^2 u \Vp}{g_s^2\pi\left(2\pi\right)^6}
\frac{Q}{\dper\Vp\tau_4\lpl u^2}
\left(\frac{\partial y}{\partial\psi}\right)^2 
=-\frac{4u^2}{\dper}=- \frac{\theta^2}{4}
\ee
For $u\simeq 1/M_s\lpl\simeq \alpha_{GUT} \simeq 1/25$ (a 
reasonable choice, where $\alpha_{GUT}$ is the standard 
model coupling at the GUT scale), the slow-roll 
condition is easily satisfied, and it is possible to obtain 
more than 60 e-foldings of inflation since $N_e\sim1/\eta$. 

A few comments are in order.
As the jellium term contribution to $\eta$ is very small, the 
number of e-foldings is dictated by the first non-zero harmonic term. 
Here it is the quartic term with strength $B_4$ given in 
Table \ref{antipodalcoef}. 
As we pointed out earlier, the $B_4$ obtained here are at least a factor 
of three smaller than those used in Ref.\cite{Jones:2002cv}. Thus the 
inter-brane potential is weaker than Ref.\cite{Jones:2002cv} uses, and so it is 
much easier to get sufficient inflation than they indicate.

The Madelung term $C_a$ will shift the 
vacuum energy of the inflaton potential. As is clear from 
Eq.(\ref{quadratic_potential}) and Table \ref{antipodalcoef}, 
this shift is positive, that is, it increases the vacuum energy term 
in the inflaton potential.  Its contribution tends to decrease 
the magnitude of $\eta$.

After collision, the two branes at a small angle reduce to two parallel 
branes (horizontal in Fig.(\ref{torus_wrapping})) with zero vacuum energy.
In an orientifold, the tension and RR charge of these two branes are 
canceled by the presence of orientifold planes. This implies that 
during inflation, the branes at angle will have a non-zero force with 
the remaining branes and orientifold planes. For example, the 
interaction between the $(1,-1)$ D4-brane and a $(1,0)$ D4-brane 
is proportional to  
\be
Q_{(1,0)} = -\frac{M_s^2}{2 \pi \sin \theta}
\left(1 -\frac{\sin^2 (\theta/2)}{2}- \cos (\theta/2) \right)
\simeq \frac{Q}{16}
\ee
and its interaction with orientifold planes is also suppressed 
by the same factor of 16, but with opposite sign.
These contribute a small correction to the interaction between the
$(1,-1)$ and the $(1,1)$ branes.
One can always place the $(1,1)$-brane, the $(1,-1)$-brane,
and the $(1,0)$-branes at initial positions that are favorable 
to inflation.  

\subsection{Other Branes-at-a-Small-Angle Inflationary Scenarios}

Next we consider branes wrapping the long dimension of the torus more 
than once.
Suppose one branes has $\left(n, 1\right)$ wrapping and the other has
 $\left(n, -1\right)$ wrapping.
After collision, we are left with $2n$ parallel branes.
In this case the angle between the branes is $\theta=2u/n$ and 
the constant piece of the potential is:
\be
V_0=2\tau_p\lpl\left(\sqrt{n^2+u^2}-n\right)\simeq\tau_p\lpl \frac{u^2}{n}
\ee
The branes now intersect in $n^2$ points, so the charge $Q$ is given by: 
\be
Q = n^2\frac{M_s^2}{2\pi}\frac{u^3}{n^3}=\frac{M_s^2}{2\pi}\frac{u^3}{n}
\ee
The relationship between $\psi$ and $y$ becomes:
\be
\psi=y\sqrt{\tau_p\lpl\frac{n}{2}}
\ee
and the slow-roll parameter $\eta$ becomes:
\be
\eta \simeq -\frac{u^2}{n}
\ee
For small $u$ and large $n$,
the slow-roll condition is easily satisfied, and it 
is possible to obtain many more than 60 e-foldings of inflation 
since $N_e \sim 1/|\eta|$. The other slow-roll parameter $\epsilon=0$
at the antipodal point. In the actual case, the slow-roll parameters 
are dictated by the quartic harmonic term where $\epsilon$ is 
negligibly small. 
For a reasonable choice $n=8$, we see that the number of e-foldings 
that can be obtained 
for inflation is further improved. In fact, the quadratic term is 
negligible in this case and the inflaton rolling is dictated by the 
quartic harmonic term discussed in Ref.\cite{Jones:2002cv}.

We may also consider branes with dimensionalities other than $p=4$.
In the $p = 6$, $\dper = 2$ case the D6-branes span the  $X^4, X^5$ torus 
and wrap different cycles of the $X^6, X^7$ torus.
They are localized in the $X^8, X^9$ torus and the interaction potential 
is logarithmic.
The potential and the Planck mass are given by:
\baray
&& M_P^2 = \frac{M_S^8V_{45}\lpl^2\theta\Vp}{g_s^2\pi\left(2\pi\right)^6} \nonumber\\
&& V\left(y\right) = \frac{\tau_{6}V_{45}\lpl\tan^2\theta}{4} - \frac{Qy^2}{2\dper\Vp} - \beta Q\log{\left(M_sy\right)}
\earay
where $Q \simeq M_s^4\theta^3/16\pi$.
Again, in the small angle approximation, the relevant slow-roll parameter becomes:
\be
\eta \simeq -\frac{\left(2\pi\right)^2\theta^2}{\left(M_s\rpl\right)^2}
\ee
The slow-roll parameter is again small in this case, and the end of 
the slow-roll is determined by the attractive logarithmic potential. 
The region yielding enough slow-roll is reasonably large and there is 
no need to fine-tune the initial conditions. 

It will be interesting to work out the situation of
other brane inflationary scenarios
\cite{Herdeiro:2001zb,Dasgupta:2002ew,Gomez-Reino:2002fs}.

\section{Discussions}

In the simplified scenario discussed here,
the $\DD$ inflationary scenario is not viable. On the other hand, 
the branes at a small angle scenario remains a viable model for 
inflation. Cosmic strings are generically produced towards the 
end of the brane inflationary epoch. Using the temperature fluctuation
in the cosmic microwave background radiation to fix the 
superstring scale, the cosmic string tension arising from the 
brane recombination in the branes at a small angle inflation 
happens to be much bigger than that in the $\DD$ scenario
\cite{Jones:2002cv,Sarangi:2002yt}. 
If branes-at-a-small-angle scenario is preferred, one consequence 
is that the cosmic string tension will be on the high side, up 
to values just below the present experimental 
bounds. This enhances the hope to test the brane inflationary
scenario via the search of signatures of cosmic strings.

As is clear from the analysis, the brane inflationary scenario 
depends on the dynamics of moduli stabilization. Presumably 
compactification moduli are stabilized by some string dynamics, 
or effective potential. The minimum of such an effective 
potential measures the cosmological constant. So understanding the
moduli stabilization problem implies some understanding of the 
cosmological constant problem, or in a less ambitious framework,
moduli stabilization must accommodate the smallness of the
observed dark energy \cite{Kachru:2003aw}. Hopefully, brane inflation 
in string theory allows us to address this important issue.

\vspace{0.3cm}

We thank Cliff Burgess, Nick Jones, Shamit Kachru, Juan Maldacena, 
Fernando Quevedo and Gary Shiu for many valuable discussions.
This material is based upon work supported by the National Science 
Foundation under Grant No.~PHY-0098631.

\appendix\section{Ewald's Method}

Here we will detail Ewald's technique and extend it to suit our purpose. Consider the $d$-dimensional torus as an infinite lattice and let $\Phi({\bf r})$ satisfy
\ba
\label{bigEw}
\nabla^2\Phi({\bf r})&=&\sum_{j}-\delta({\bf \Delta r_j}) + \frac{1}{\Vol}\\\nonumber
\Phi({\bf r})&=&\Phi_1({\bf r})+\Phi_2({\bf r}) + constant \\\nonumber
\nabla^2\Phi_1({\bf r})&=& \sum_{j}-P(\Delta r_j)\exp(-\epsilon^2\Delta r_j^2) + 1/\Vol \\\nonumber
\nabla^2\Phi_2({\bf r}) &=& \sum_j\left[-\delta\left({\bf \Delta r_j}\right)+ P(\Delta r_j)\exp(-\epsilon^2\Delta r_j^2)\right]
\ea
Where ${\bf \Delta r_j}={\bf r-r_j}$, and ${\bf r_j}$ is the $j^{th}$ lattice vector. Here $P(\Delta r_j)\exp(-\epsilon^2\Delta r_j^2)$ can be thought of as a charge distribution inserted at the $j^{th}$ lattice site, normalized so that
\be
\int_{{\mathbbm R}^d} P(r)\exp(-\epsilon^2r^2)dv = 1
\ee
The charge distribution may extend outside the unit cell, but integrating one distribution over all space is identical to integrating all distributions over the unit cell. 
The effect of adding and subtracting a diffuse charge distribution is to regularize the summation over the lattice. The full potential $\Phi({\bf r})$ is the sum of $\Phi_2({\bf r})$ (the real-space sum) and $\Phi_1({\bf r})$ (the reciprocal lattice sum). For a good choice of $P(r)$ the sums are separately convergent, and together reproduce $\Phi({\bf r})$ for the original point-charge distribution. Clearly, $\Phi({\bf r})$ should be independent of $P(r)$ and $\epsilon$, which are chosen to enhance the convergence of the sums. There is a range of choices of $\epsilon$ (for our calculation $18\leq\epsilon\leq24$) for which the results of Eq.(\ref{bigEw}) are independent of $\epsilon$. Other crucial conditions are listed and demonstrated in the next section. This procedure is well tested since it is widely used in solid state physics to compute the Madelung constant.

\subsection{Three dimensions}
In three dimensions, a simple choice is to add and subtract a Gaussian charge distribution at each lattice site. This gives $P(r)=\epsilon^3\pi^{-3/2}$. Then we have
\ba
\label{3d}
\Phi_{1}^{(3)}({\bf r})&=& \frac{1}{\Vol}\sum_{{\bf k}_j\neq {\bf 0}}
\frac{\exp(-k_{j}^2/4\epsilon^2)\exp(i{\bf k}_{j}\cdot\bf{r})}
{k_j^2} \\\nonumber
\Phi_{2}^{(3)}({\bf r})&=& \frac{1}{4\pi}\sum_{\Delta r_j}\frac{1-{\rm erf}(\epsilon \Delta r_j)}{\Delta r_j}
\ea
where $\bf {k}_j$ is the $j^{th}$ reciprocal lattice vector. Note that the following necessary conditions are satisfied:\\
1) The terms in $\Phi_1$ converge faster than $k_j^{-3}$.\\
2) $\Phi_1$ reduces to Eq(\ref{Ewald}) when $\epsilon\rightarrow\infty$.\\ 
3) The terms in $\Phi_2$ converge faster than $\Delta r_j^{-3}$.\\ 
4) $\Phi_2$ goes to zero as $\epsilon\rightarrow\infty$.\\ 
5) $\Phi_2$ reduces to $\sum_j\Delta r_j^{-1}$ as $\Delta r_j\rightarrow0$.\\ 
Any other choice for $P(r)$ must lead to potentials that satisfy these same limits for the method to work.

In principle there is also an arbitrary constant of integration, but 
the requirement that $\Phi$ be independent of $\epsilon$ yields
\be
\label{phi3d}
\Phi^{(3)}({\bf r})=\Phi_{1}^{(3)}({\bf r}) + \Phi_{2}^{(3)}({\bf r}) -\frac{1}{4\Vol\epsilon^2}
\ee
The $\epsilon$-dependent term comes from the missing zero mode in $\Phi_1(r)$, as can be seen by differentiating with respect to $\epsilon$. The constant vanishes as $\epsilon\rightarrow\infty$, so that Eq.(\ref{phi3d}) recovers Eq.(\ref{Ewald}). This also satisfies Eq.(\ref{intGz}). In practice, a suitable choice of $\epsilon$ increases the convergence of the calculation. In condensed matter a jellium is always used to make the lattice neutral, but a more general term can be easily incorporated into this technique by including the appropriate Fourier expansion in $\Phi_1$. 

The numerical potential generated by summing Eq.(\ref {phi3d}) over the lattice can be fit to a potential of the form given in Eq.(\ref {genform})
\ba
\Phi^{(3)}({\bf r})&=& \frac{1}{4\pi r} + \frac{r^2}{6\Vol}+ \frac{C^{(3)}_s}{\Vol^{\frac{1}{3}}} + A^{(3)}_4h^{(3)}_4(x_i) + A^{(3)}_6h^{(3)}_6(x_i)+\cdots
\ea
where the quadratic piece comes from the jellium, and the constant is related to the Madelung energy of the lattice. The data generated by the Ewald method can also be used to fit the potential near the antipodal point. We expect this expansion to be harmonic, except for the $r^2$ term. With coordinates now measured from the center of each cube, this is
\ba
\label{3dantipodal}
\Phi^{(3)}_{antipodal}({\bf z})& = & \frac{C^{(3)}_{a}}{\Vol^{\frac{1}{3}}} + \frac{1}{6\Vol}\sum_{i=1}^{3}z_i^2 
+ B^{(3)}_4h^{(3)}_4(z_i)+B_6^{(3)}h^{(3)}_6(z_i) + \cdots
\ea
Of course, if the expression near the podal point is known exactly, Eq(\ref{3dantipodal}) can be obtained algebraically using $z_1=x_1-\frac{L}{2}$, etc. Generally, though, the higher order coefficients $A_{10}, A_{12},\dots$ will not be small, so it is best to use a numerical fit. 

\subsection{Four dimensions}
To implement the same procedure in dimensions other than three, we first need to choose a suitable form for $P(r)$. We will need the Laplacian in $d$ dimensions:
\be
\nabla^2\varphi(r)=\frac{1}{r^{d-1}}\frac{\partial}{\partial r}\left[r^{d-1}\frac{\partial}{\partial r}\varphi\right]
\ee
The choice for $P(r)$ is guided by the behavior of $\Phi_1$ and $\Phi_2$ in the limits mentioned below Eq(\ref{3d}). In particular, each should converge at least faster than $\frac{1}{r^{d-2}}$ or $\frac{1}{k^{d-2}}$ since the number of points in the sums over the lattice grow as the ($d-1$) power of $r$ or $k$. Also, $\Phi_2({\bf r})$ should reduce to the usual Coloumb potential as $r\rightarrow0$. Taking $\epsilon\rightarrow\infty$, which amounts to eliminating the added charge distributions, recovers Eq(\ref{Ewald}). In four dimensions, a simple choice is \cite{ewald}:
\be
\Phi_2^{(4)}({\bf r})=\frac{1}{4\pi^2}\sum_{\Delta r_j}\exp(-\epsilon^2\Delta r_j^2)\left(\frac{1}{\Delta r_j^2}+a\epsilon^2+b\epsilon^4\Delta r_j^2+...\right)
\ee
The coefficients ($a,b,..$) can be adjusted to give a single term in the charge distribution $P(r)\propto r^n$. The choice of how many higher powers of $r$ to include in the above equation is a matter of taste and simply changes $n$. Consider $a=\frac{1}{2}, b=0$. Then 
\ba
P^{(4)}(r)&=&\frac{\epsilon^6r^2}{2\pi^2}\\\nonumber
\Phi_{1}^{(4)}({\bf r})&=& \frac{1}{\Vol}\sum_{{\bf k}_j\neq {\bf 0}}
\frac{\exp(-k_{j}^2/4\epsilon^2)\exp(i{\bf k}_{j}\cdot\bf{r})}
{k_j^2}\left(1-\frac{k_{j}^2}{8\epsilon^2}\right)\\\nonumber
\Phi^{(4)}({\bf r})&=&\Phi_1^{(4)}({\bf r})+\Phi_2^{(4)}({\bf r})-\frac{3}{8\Vol\epsilon^2}
\ea
The constant in the last expression above guarantees that $\Phi$ is independent of $\epsilon$. 

\subsection{Two dimensions}
In 2 dimensions, a useful form for the real-space part is
\be
\Phi^{(2)}_2({\bf r})=\frac{1}{2\pi}\sum_{\Delta r_j}\int_{\Delta r_j}^{\infty}\exp(-\epsilon^2r'^2)\left(\frac{1}{r'}+a\epsilon^2r' +b\epsilon^4r'^3+...\right)dr'
\ee
To evaluate $\Phi_2$ in the Ewald sum, note that some convienent expressions for the exponential integral are:
\ba
\int_{t}^{\infty}\frac{\exp(-u)}{u}du &=&
-\gamma_E-{\rm log(t)}-\sum_{n=1}^{\infty}\frac{(-1)^nt^n}{n\cdot n!}\\
&=& \exp(-t)\frac{1}{1+t-\frac{1}{3+t-\frac{4}{5+t-...}}}
\ea
For the simplest choice $a=b=0$, we have
\ba
P^{(2)}(r)&=&\frac{\epsilon^2}{\pi}\\\nonumber
\Phi_1^{(2)}({\bf r})&=&\frac{1}{\Vol}\sum_{{\bf k}_j\neq {\bf 0}}
\frac{\exp(-k_{j}^2/4\epsilon^2)\exp(i{\bf k}_{j}\cdot\bf{r})}
{k_j^2}\\\nonumber
\Phi^{(2)}({\bf r})&=&\Phi_1^{(2)}({\bf r})+\Phi_2^{(2)}({\bf r})-\frac{1}{4\Vol\epsilon^2}
\ea
Extending the techniques from the previous three sections, we now have a well-defined way to evaluate the potential numerically for general dimension, charge configuration, and metric.

\end{document}